# Temperature Dependence of Electric Transport in Few-layer Graphene under Large Charge Doping Induced by Electrochemical Gating

R. S. Gonnelli[1], F. Paolucci[1(♣)], E. Piatti[1], Kanudha Sharda[1], A. Sola[1(♦)], M. Tortello[1], Jijeesh R. Nair[1], C. Gerbaldi[1], M. Bruna[2(♥)] and S. Borini[2(♥)]

[1] *Dipartimento di Scienza Applicata e Tecnologia, Politecnico di Torino, 10129 Torino, Italy*
[2] *Istituto Nazionale di Ricerca Metrologica (INRIM), Torino, 10135 Torino*

(♣) *Present address: Max Planck Institute for Solid State Research, Stuttgart, Germany*
(♦) *Present address: Istituto Nazionale di Ricerca Metrologica (INRIM), Torino, 10135 Torino*
(♥) *Present address: Nokia Research Center, Cambridge, CB3 0FA, United Kingdom*

**The temperature dependence of electric transport properties of single-layer and few-layer graphene at large charge doping is of great interest both for the study of the scattering processes dominating the conductivity at different temperatures and in view of the theoretically predicted possibility to reach the superconducting state in such extreme conditions. Here we present the results obtained in 3-, 4- and 5-layer graphene devices down to 3.5 K, where a large surface charge density up to about $6.8 \cdot 10^{14}$ cm$^{-2}$ has been reached by employing a novel polymer electrolyte solution for the electrochemical gating. In contrast with recent results obtained in single-layer graphene, the temperature dependence of the sheet resistance between 20 K and 280 K shows a low-temperature dominance of a $T^2$ component – that can be associated with electron-electron scattering – and, at about 100 K, a crossover to the classic electron-phonon regime. Unexpectedly this crossover does not show any dependence on the induced charge density, i.e. on the large tuning of the Fermi energy.**

**Introduction**

The electric transport properties of single-layer graphene (SLG) have been the subject of a very extensive experimental and theoretical research activity in the past ten years after the breakthrough discovery of the possibility to isolate and characterize this 2D material [1, 2]. A comprehensive review paper covering the first six years of this activity can be found in literature [3], while some additional information is also present in [4]. Although the transport properties of multi-layer graphene have been experimentally studied [5, 6], the electric transport of few-layer graphene (FLG, with number of layers $N$ between 2 and 5) has been much less intensively investigated [7]. This is particularly true when one considers the low-temperature behavior of these transport properties. On the other hand, this information can provide important clues on the scattering processes dominating the conductivity of FLG and on the quantum corrections to the Drude conductivity which may occur at low temperature [3,4]. Moreover, the possibility to strongly change the number of charge carriers by doping the FLG above the limits of the standard solid-dielectric backgate technique [1, 2, 8, 9] adds an additional degree of freedom to the study of the electron scattering in these materials. But, in our opinion, there is another important reason for performing low-temperature high-doping transport experiments in FLG, related to the theoretically predicted possibility of the appearance of a superconducting state in highly-doped or metal (K, Li) decorated SLG [10−16] or in hole-doped graphane [17], which is a reversibly idrogenated graphene [18]. In particular, a conventional BCS phonon-mediated superconductivity at about 8 K has been predicted in Li-covered SLG [13] with a $T_c$ that can increase up to 17-18 K in case of lithium coverage on both sides of graphene [13] or up to 13.5 K in case of a $C_6$-Li-$C_6$-Li layered structure [14]. Another proposed route for the possible obtainment of an unconventional chiral superconductivity in graphene [15] or a more conventional two-gap BCS superconductivity with $T_c$ up to about 13 K [16] relies on the very strong (n-)doping of SLG. Due to the very large interest in the possible applications of graphene in electronics, the observation of superconductivity (even at low T) in this material would obviously lead to new nanoscale device concepts or to the performance improvement of the existing ones.

Independently of the different approaches and the different results, in all the mentioned papers the key ingredients for the obtainment of superconductivity appear to be: i) a large increase of the carrier density, able to shift the Fermi level far from the Dirac point and close to a Van-Hove singularity (VHS); ii) the presence of a parabolic interlayer-like band that is produced by the deposited metal adatoms [10, 13, 14] or by the large doping itself [12, 15, 16]. We believe that both these ingredients could be also present in the few-layer graphene stacks at a very high level of induced charge density.

Interestingly, double multi-layer graphene structures separated by an insulator have been proposed to realize strong-coupling electron-hole superfluidity (leading to counter flow superconductivity) [19]. In addition, VHSs in double 3- and 4-layer graphene devices have been exploited to enhance critical temperature and onset density of the electron-hole superfluid transition [20].

For this reason, even in absence of precise model predictions for superconductivity in FLG, we thought it was worthwhile to perform electric transport measurements in FLG at low temperature and high doping.

In the past few years the possibility to dope various materials at an unprecedented level of surface charge density – even inducing superconductivity in materials initially insulating – has been proven [21, 22]. It relies on the use of a polymer electrolyte or a liquid gating technique. In practice an electrochemical cell is incorporated into a top-gate field-effect transistor (FET) architecture replacing the solid gate insulator by a polymer electrolyte solution (PES) or a liquid electrolyte that acts as gate dielectric. A voltage application between a gate electrode and another electrode in contact with the material under study causes solvated ions in the PES to densely accumulate on the surface of a channel over the sample, thus forming an electric double layer (EDL) that acts as a nanoscale capacitor at the interface. As a consequence of the very large specific geometric capacitance of the EDL (of the order of some $\mu F \cdot cm^{-2}$) an intense electric field as high as 50 MV/cm can be applied at the sample surface inducing a huge surface charge density $n_{2D}$, up to $10^{14}$-$10^{15}$ charges/cm$^2$ depending on the electronic properties of the material [22 and references therein]. This technique has been recently used in Raman and IR conductivity experiments in SLG and bilayer graphene up to induced charges of the order of $5 \cdot 10^{13}$ cm$^{-2}$ [23-26]. However not so many papers are present in literature concerning the application of this technique to transport measurements in SLG or FLG. In a first experiment in SLG a charge induction up to $\sim 6 \cdot 10^{13}$ cm$^{-2}$ has been demonstrated by using polyethylene oxide (PEO) as polymer and lithium perchlorate as electrolyte [27]. This large surface density has been further increased up to $n_{2D} > 10^{14}$ cm$^{-2}$ by using the same PES at gate voltages up to 15 V, but employing a rapid cooling method in order to reduce the expected electrochemical reactions between PES and graphene [28]. The authors found a quartic temperature dependence of the resistivity at low T (as expected for scattering with 2D phonon modes) and a crossover to the classic high-T linear behavior at a temperature strongly tunable by the induced charge [28]. More recently the technique has been also applied to FLG. The Dirac curves of single, bi, and trilayer graphene have been compared at room T and deeply analyzed up to charge densities of the order of $2 \cdot 10^{14}$ cm$^{-2}$ [29] and the low-temperature magnetoconductance of epitaxial trilayer graphene grown on a 6H-SiC surface has been studied up to n $\sim 7 \cdot 10^{14}$ cm$^{-2}$ claiming the evidence for a gate modulation of the weak spin-orbit interaction in this material [30]. No sign of superconductivity has been observed in the low-temperature sheet resistance curves shown in these articles [27, 28, 30].

In this paper we present the results of electric transport measurements in 3-, 4- and 5-layer exfoliated graphene samples (3LG, 4LG and 5LG in the following) down to 3.5 K and up to an induced surface charge density that ranges from about $5 \cdot 10^{14}$ cm$^{-2}$ in 3LG to about $7 \cdot 10^{14}$ cm$^{-2}$ in 5LG obtained by using a novel PES of improved efficiency. No trace of superconducting transition has been found in these experiments, but the sheet resistances in the temperature range 20 – 280 K reproducibly show a peculiar behavior with a dominant low-temperature T$^2$ component and a crossover to the classic electron-phonon linear regime at about 90-100 K, independently of the strong Fermi energy tuning produced by the electrochemical gating.

**Results and Discussion**

The FLG flakes used in this work have been obtained by using the standard adhesive tape exfoliation technique followed by a transfer onto a SiO$_2$ on Si substrate. The number of graphene layers has been determined by optical contrast [31] and by Raman spectroscopy measurements. The latter technique was also used for selecting only the 3LG, 4LG and 5LG flakes that showed a Bernal stacking [32, 33]. The electric contacts for the four-wire measurements were realized by using the standard photolithographic, thermal evaporation and lift-off techniques. Afterwards, the FLG channel was shaped into a Hall-bar by reactive ion etching and an additional windowing by photoresist was realized on the device in order to leave uncovered only a part of the contact pads (for wire bonding) and the gated channel over the flake. Figure 1a shows a picture of one of the devices used in our experiments, taken under an optical microscope. In the inset a zoom in the region of the contacts and of the FLG Hall-bar-shaped flake is presented, where the detail of the photoresist window defining the channel geometry can be seen.

The EDL gating was realized by using a novel PES that proved to be able to induce record surface charges up to $4.5 \cdot 10^{15}$ cm$^{-2}$ in previous experiments we performed on thin films of Au, Cu and Ag [34, 35]. In Figure 1b a schematic view of the device with the FLG channel covered by the PES, the Pt wire used as gate electrode and the various connections used for longitudinal ($R_{xx}$) and Hall ($R_{xy}$) resistance measurements are shown. Before cooling down the device the ambipolar nature of the field effect in FLG was always checked by performing sheet conductance measurements during slow gate voltage sweeps between 3 and -3 V or vice-versa. One of these sheet conductance vs. gate voltage $V_G$ curves is reported in Figure 1c for a 4LG device. Up to $V_G \sim \pm 1$ V (with respect to the neutrality point) the curve is symmetric, but at a higher gate voltage a pronounced asymmetry is observed, together with a broad depression at about $V_G = 1.5$ V. Similar features have been observed in 3LG where the depression has been related to the crossing of a VHS when the Fermi level reaches the split band $T_{2g}$ [29].

The determination of the induced charge density is particularly important in these experiments. We used three different methods for accessing this quantity. The first one consists in the classic Hall-effect measurements that we performed at room temperature by using a calibrated variable-gap magnet made with permanent magnets and able to apply magnetic fields up to 0.7 T at the sample surface. The second method is a typical electrochemical technique well described in literature [36] and known as double-step chronocoulometry (DSCC). It consists in the successive application and removal of the gate voltage in the form of a step-wise function while measuring the small current (of the order of few nA) that flows between the gate electrode and the device (gate current $I_G$). The time dependence of this current shows a fast exponential component connected to the formation of the EDL and a slower minor one that pertains to electrochemical reactions possibly occurring between the PES and the sample. The inset of Fig. 2 shows a typical time dependence of $I_G$ under removal of a 3 V gate voltage in a 4LG device. By fitting with the proper electrochemical model [36] the $I_G$ vs. time curves it is possible to determine the charge that forms the double layer and, dividing by the gated area, to obtain the induced surface charge density $n_{2D}$ [34, 35]. The main panel of Fig. 2 shows a comparison between the $n_{2D}$ values determined by Hall effect measurements in 4LG (orange diamonds) and the ones obtained from DSCC (black, red and blue circles). The induced charge density was measured in 3LG (black circles) and then the values for 4LG and 5LG have been obtained by scaling the 3LG results with the calculated ratio of their quantum capacitance $C_q$ [37].

A further consistency check of the results shown in Fig. 2 is possible. In fact, it is well known that in case of gating of a 2D material the EDL plus the 2D electron-liquid system can be modeled by a series of two specific capacitances: the geometric one $C_g$, which describes the EDL, and the quantum one, $C_q$, that essentially describes the screening properties of the 2D conductor. As a consequence $1/C_{EDL} = 1/C_g + 1/C_q$. The geometrical capacitance $C_g$ depends on the properties of the PES and on $V_G$ and can be determined from our gating experiments in Au thin films [34] where, of course, $C_q \rightarrow \infty$. At $V_G = 4$V it turns out to be in the range 50-140 μF/cm$^2$ depending on the "freshness" of the PES [35]. The quantum capacitance $C_q$ of our devices can be estimated from tight-binding and ab-initio DFT calculations (jellium model) of the effective electron mass $m^*$ of the FLG flakes at the different charge dopings [38] by using the original definition of $C_q$ derived by Luryi [39]:

$$C_q = \frac{g_v m^* e^2}{\pi \hbar^2}$$

where $g_v$ is the valley degeneracy factor. At $V_G = 4$V, for example, we estimate $C_q = 24 \pm 7$ and $27 \pm 8$ μF/cm$^2$ for 4LG and 5LG, respectively, in good agreement with the values measured in 3LG at high charge doping [29]. From the above values of $C_g$ and $C_q$ we can calculate $C_{EDL}$ and, consequently, $n_{2D}$ for 4LG and 5LG at $V_G = 4$V. The obtained ranges of values (including only the variability of $C_g$) are shown in Fig. 2 as dashed bands (red for 4LG and blue for 5LG). By comparing the carrier density determined by Hall measurements with that obtained by DSCC it is possible to notice that the Hall values are somewhat larger especially at larger gate voltages. We think that the reason for this behavior is related to the low values of the magnetic field we used in the Hall measurements. There are strong non-linearities in the Hall voltage at low field mainly due to inhomogeneities and differences in the mobility of the sample, that introduce a large uncertainty in the measure (see error bars in Fig. 2). More precise Hall values of the average charge density can be obtained at fields of the order of 5 T that are not available in our cryo-cooler experimental setup. In the range where the comparison of the three methods of $n_{2D}$ determination is possible ($V_G \leq 2$V) the average $n_{2D}$ value coincides with the one obtained from DSCC and thus we conservatively decided to use these DSCC values in the whole $V_G$ range including the proper error (of the order of $\pm 30\%$ at high doping) in the $n_{2D}$ evaluation. Please note that, even if the redox potential of Li$^+$ is 3.04 V, we have experimentally observed by long-term monitoring the device resistance and the gate current at room temperature that a $V_G$ up to +4 V can be safely applied without any evidence of electrochemical reactions at the interface.

Figure 3 shows the sheet resistances as a function of temperature obtained at different induced charge densities (indicated in the legends) in 3LG, 4LG and 5LG devices. The curves show some common features. The sheet resistances $R_\square$ of 3LG and 4LG devices (panels a, b and c) exhibit a metallic behavior even at the lowest doping ($V_G = 0$ and $n_{2D}$ slightly negative), while the $R_\square$ of 5LG in the same conditions (panel d) is increasing at the decrease of temperature showing a typical localization-regime behavior. This particular behavior of the 5LG device needs to be fully clarified, but it could be due to the presence of a large number of defects and impurities at the channel surface able to introduce a strong scattering responsible for the localization at low $n_{2D}$. However, the increase of the charge density (both positive and negative) produces a strong decrease of the sheet resistances and, in the 5LG device, leads to a crossover from localization to a metallic regime at $V_G > 1$ V, i.e. $n_{2D} > 4.3 \cdot 10^{13}$ cm$^{-2}$. A similar behavior has been recently observed in liquid-gated 3LG devices [30].

No sign of superconducting transition can be seen at the lowest temperature reached in the experiments, i.e. 3.5 K, but a logarithmic upturn of the sheet resistance is present at $T < 15$ K in all the FLG devices. This behavior will be analyzed and discussed in detail in a forthcoming paper. Let us focus here on the range of higher temperatures, i.e. when T > 20 K. When the resistances are in the complete metallic regime (any $V_G$ for 3LG and 4LG, $V_G \geq 2$V, i.e. $n_{2D} \geq 1.85 \cdot 10^{14}$ cm$^{-2}$ for 5LG) they show a linear behavior as a function of temperature in the range between about 100 K and 270 K. The slope of this high-temperature linear-T dependence practically does not change at the increase of the charge density $n_{2D}$ (both positive and negative) as it can be seen from the curves of Fig. 3 and has been already observed in SLG [8, 28]. Below ~ 100 K the curves change behavior assuming a steeper T dependence that is appreciable in the curves of Fig. 3a, b and c, but is also present (even if not visible due to the different scale) in the curves of Fig. 3d.

The standard way to get more information on the temperature dependence of the resistance curves is to plot them in a double logarithmic scale. Figure 4 shows an example of these log $(R_\square - R_{\square\,min})$ vs. log T curves in the case of the 4LG device and in the temperature range 20-280 K for both positive (panel a) and negative (panel b) charge induction. Here $R_{\square\,min}$ is the minimum of the sheet resistance (at $T \sim 0$ K) obtained by fitting the temperature dependency in the range 20-90 K (see below). In both cases a linear $T$ dependence is observed at $T \gtrsim 100$ K. Below ~ 80 K the temperature dependence becomes steeper, showing a dominant $T^2$ component (see Fig. 4b). In the intermediate temperature range (70 K $\leq T \leq$ 100 K) we see a rather sharp (at least for positive gating, i.e. electron doping) crossover between the two regimes. These results are confirmed by the temperature dependence of the carrier mobility that is shown in Fig. 5 for the 4LG device both for positive (panel a) and negative (panel b) charge induction. Also in this graph the crossover around 100 K toward a steeper low-temperature dependence and the progressive reduction of the slope of the mobility below 30-40 K can be seen in the curves at lower charge doping. These results have been reproducibly observed also in the 3LG and 5LG devices.

A "classic" small-angle electron-phonon scattering should dominate the sheet resistance below the Bloch-Grüneisen temperature $\Theta_{BG}$ in a low-density electron system. This should lead to the well-known $T^5$ dependence in the 3D case and to a $T^4$ dependence in the 2D one, as theoretically predicted [40] and experimentally observed in SLG [28]. However, in our case, it is clearly impossible to fit the experimental data with a $T^4$ dependence (shown, as reference, in both the panels of Fig. 4). As a consequence, this weaker fall-off of the sheet resistance with decreasing $T$ in the gated few-layer graphene samples compared to single-layer-graphene (Ref. 28) is certainly somewhat unexpected, even if previously observed and discussed in SLG devices [41]. As we have already pointed out, the geometric capacitance of the EDL strongly depends on the composition of the PES and its preparation and, therefore, also the carrier transport in the channel could be sensitive to this composition. One can argue that the difference in the low-temperature behavior of SLG and FLG is due to the different PES used in the two experiments. This is not the case. In Fig. 5c we show the log-log plot of the temperature dependence of the sheet resistance measured in a 3LG and in a 4LG device before drop casting the PES. It is quite evident that the $T^2$ dependence and the crossover to the linear-$T$ behavior at about 100 K are present in our devices even when the PES is not put over the channel. In our experiments we always observed that, apart from a small change in the charge density (indicated by a shift in the sheet resistance), the presence of the PES (at $V_G$ =0) does not generate any "discontinuity" in the behavior of the resistance curves as a function of temperature. As a consequence the peculiar $T^2$ dependence of these FLG devices cannot be attributed to the gating and to the large carrier density it introduces, but is likely due to the multiband nature of their electronic spectrum or, less likely, to extrinsic effects related to the fabrication of the devices.

The most straightforward explanation for the $T^2$ dependence of the resistance below ~ 80 K that progressively turns to a $T$ dependence under 25-30 K (see Fig. 4a and, especially, Fig. 5c) is that in our devices the intermediate- and low-temperature sheet resistance is dominated by electron-electron collisions with both small momentum transfer (Nyquist term, $R_\square \propto T$) and with large momentum transfer (direct ballistic term, $R_\square \propto T^2$). A very similar low-$T$ behavior has been already observed in SLG, but not discussed at $T > 60$ K [41]. In order to quantitatively confirm this observation we fitted the sheet resistances of *all* our devices in the temperature range 20-90 K by the function: $R_{\square\,min} + aT + bT^2$ where $R_{\square\,min}$, a and b are fitting parameters and the quadratic term is always more or less dominant at $T \gtrsim 30$ K. Figure 6a shows two very good examples of these best fits (black dashed lines) in the case of the 4LG device at two different gating levels, $V_G = 3$ V (green continuous line) and 4 V (orange continuous line). The weight of the quadratic component varies with the number of graphene layers and the sign of doping, but, in any case, it becomes small at $T < 20$-30 K and the Nyquist term finally becomes dominant at lower temperatures.

As far as the crossover to the high-temperature linear $T$ dependence is concerned we can simply note that, independently from the value and the sign of the induced charge density (for example from $0.39 \cdot 10^{14}$ cm$^{-2}$ to $6.19 \cdot 10^{14}$ cm$^{-2}$ as shown in Fig. 4a), it always occurs at a temperature of 90-120 K. This temperature is very close to the one where a similar crossover (but from a $T^4$ to a $T$ behavior) has been observed in SLG at $n_{2D} > 4.65 \cdot 10^{13}$ cm$^{-2}$ (see Fig. 2 of Ref. 28). This temperature cannot be associated to the Debye temperature $\Theta_D$ of the material since it should be higher than 2000 K ($\Theta_D \approx 2300$ K in SLG [28] and $\Theta_D \approx 2500$ K for in-plane phonon modes of graphite) thus potentially leading to a crossover at $T > 400$ K ($\sim 0.2\ \Theta_D$).

Following Ref. 28 it is thus tempting to interpret also in our case this crossover temperature as related to the Bloch-Grüneisen temperature $\Theta_{BG}$ of the FLGs at our high doping levels. In a simple, single-band and free-electron-like picture $\Theta_{BG}$ is the temperature at which the maximum momentum of acoustic phonons equals twice the Fermi momentum and is therefore able to completely span the Fermi surface. In this ideal case $\Theta_{BG}$ is given by:

$$\Theta_{BG} = \frac{2\hbar k_F v_s}{k_B} \qquad (2)$$

where $k_F$ is the Fermi wave vector of the spherical (or cylindrical) Fermi surface and $v_s$ is the sound velocity. Of course we can anticipate right now that this simplified expression will not be adequate to describe our FLG multiband devices.

We fitted our sheet resistance curves by using a generalized Bloch-Grüneisen model similar to the one introduced in Ref. 40, but with the low-$T$ exponent left as a parameter, i.e. by the function:

$$\Delta R_\square(T) = R_\square(T) - R_{\square\,min} = A \cdot \int_0^1 \left( \frac{\Theta_{BG}}{T} x^m \sqrt{1-x^2} e^{\frac{\Theta_{BG}}{T}x} \right) \Big/ \left( e^{\frac{\Theta_{BG}}{T}x} - 1 \right)^2 dx \qquad (3)$$

where $A, m$ and $\Theta_{BG}$ are free parameters. Two examples of these fits, corresponding to the electron densities $n_{2D} = 1.73 \cdot 10^{14}$ cm$^{-2}$ and $n_{2D} = -1.75 \cdot 10^{14}$ cm$^{-2}$ are shown as dash-dot curves in Fig. 4a and 4b, respectively. In the fits of all the curves of Fig. 4a and b (not shown here for clarity) the $m$ parameter is 1.85 for electron doping (Fig. 4a) and ranges between 2.1 and 2.5 for hole doping (Fig. 4b), while $\Theta_{BG}$ values range between 350 and 400 K for electron doping and between 350 and 430 K for hole doping. In Fig. 6b the parameters $\Theta_{BG}$ and $m$ obtained from the fit of the resistances of 4LG and 5LG devices using the model of eq. (3) are plotted as function of the charge density $n_{2D}$.

This figure clearly show that, apart from an increase of about 100 K of the average value passing from 4LG to 5LG, $\Theta_{BG}$ doesn't show any specific trend as a function of $n_{2D}$.

Summarizing up to now, the comparison of our present results with those previously reported in literature leads to the following conclusions: i) in 3LG, 4LG and 5LG we were able to induce a charge density quite higher than that observed in SLG and 2LG [27-29] and similar to the one recently obtained by liquid gating in 3LG [30]; ii) as in the few previous low-temperature experiments in SLG [27, 28] and 3LG [30] no superconducting transition has been observed down to ~ 3.5 K; iii) in contrast with some of the results in SLG [28], but in agreement with others [41] the dominant scattering mechanism in the range 20-90 K appears to be the direct electron-electron scattering (quadratic dependence of the temperature behavior of the sheet resistance); iv) in contrast with what theoretically predicted [40] and measured in SLG [28] the observed crossover from the low-temperature $T^2$ behavior to the standard linear-T dependence at $T \gtrsim 100$ K (and the related $\Theta_{BG}$ obtained by the fit of the sheet resistance using Eq. 3) is independent of the doping up to the maximum induced charge in 3LG, 4LG and 5LG.

At a first and superficial analysis in the framework of Eq. (2), since any $k_F$ in FLGs increases at the increase of doping (even if not proportionally to $\sqrt{n_{2D}}$ as in SLG), the absence of $\Theta_{BG}$ tuning with the Fermi energy could be related to a reduction of the sound velocity produced by the strong charge doping. As a matter of fact a similar effect of tuning of the phonon dispersion relations with a decrease of the sound velocity of some acoustic modes (and consequent softening of the corresponding part of the phonon spectrum) has been already predicted in SLG at a very high level of charge doping [16]. Even though this effect could be present in FLGs, the situation here is quite more complex. First of all the FLG flakes with Bernal stacking are always a multiband electron system. For example, even at low electron doping (when $E_F$ is tens of meV above the neutrality point) the 4LG is a two-band material with a Fermi surface made of two sheets and with two Fermi wave vectors that strongly depend on the direction in the $k$ space. The situation becomes even more complex at the increase of the electron charge density. When $E_F \sim 230$ meV a third band crosses the Fermi level and a fourth does the same at $E_F \sim 600$ meV. These Fermi energy shifts are certainly compatible with the large electron densities obtained with our PES gating. In addition, in the presence of several bands, not only intraband scattering processes but also interband ones are possible, thus considerably complicating the picture. As a consequence, in contrast with the case of SLG, in FLGs Eq. (2) cannot be used anymore as a definition of $\Theta_{BG}$. In this case, the constancy of the crossover temperature might arise from the interplay between the increase of the number and of the overall size of the FLG Fermi surfaces (at the increase of electron doping) and the presence of interband scattering processes. A quite similar situation can occur in the presence of a large hole doping. Only first-principles DFT calculations of the electron-phonon interaction accompanied by a semi-analytical solution of the Boltzmann equation in FLGs (as the one recently appeared in literature for SLG [42]) will clarify the causes of this crossover invariance.

In conclusion, by gating with a novel PES of enhanced charge-induction capability, we were able to dope 3-, 4- and 5-layer graphene devices at charge density levels that, we believe, are close to the intrinsic maxima fixed by the quantum capacitance of the devices, and to measure their sheet resistance down to 3.5 K. No traces of superconducting transition have been observed, but the temperature dependence of the resistance showed a crossover from a low-$T$ regime dominated by electron-electron scattering and a regime at higher $T$ ($\gtrsim 100$ K) where only the standard high-temperature electron-phonon scattering exists. The crossover temperature can be associated to the Bloch-Grüneisen temperature of the material, but it does not show any dependence on the Fermi energy, pointing to important differences in the electric transport properties of FLGs and SLG at the large charge densities induced by PES gating. Further ab-initio theoretical calculations of the electron and phonon properties of these materials in presence of a huge induced charge density, as well as further experiments under PES gating able to provide direct information on these properties, could shed light on these differences and their origin.

## Methods

**Device fabrication.** Few layers graphene (FLG) flakes are deposited on a 285 nm thick $SiO_2$ on Si substrates by adhesive tape exfoliation of natural graphite. Then, the samples are analyzed by optical microscopy [31], in order to estimate the number of graphene layers composing the deposited thin flake and by Raman spectroscopy for the determination of their stacking sequence [32, 33]. Only FLG flakes with Bernal stacking have been used in our transport measurements.
The geometry of the contacts is defined using photolithography followed by Cr/Au (5 nm/60 nm) thermal evaporation and lift-off. The Hall bar geometry is defined by creating a photoresist (PR) mask by photolithography and etching the uncovered portion of the film by $O_2$/Ar reactive ion etching. A further layer of PR is then spun on the sample. Windows are open by photolithography in this layer only on the FLG channel and on the contact pads for device wire bonding and mounting (see Fig. 1a). The PR mask is then hard-baked at 145 $^o$C for 5 minutes in order to improve its chemical stability.

**PES composition and preparation.** The reactive liquid formulation for the preparation of PES is made of a dimethacrylate oligomer, i.e. bisphenol A ethoxylatedimethacrylate (average $M_w \sim$ 1700 daltons, Sigma Aldrich) and a mono methacrylate based reactive diluent, i.e. poly(ethylene glycol) monomethyl ether monomethacrylate (average $M_w \sim$500 daltons, Sigma Aldrich) in 7:3 ratio along with 10 wt % of lithium bistrifluoromethanesulfonimidate salt (LiTFSI) and 3 wt % of free radical photo initiator (Darocur 1173). The components are thoroughly mixed at ambient temperature to obtain a homogeneous transparent viscous solution which is drop casted over the device [43] with the help of an optical microscope. Once the deposition is done, the whole device is exposed to UV light (light intensity 35 mWcm$^{-2}$). Under UV exposure, the initiator decomposes into free radicals and reacts with the monomers and oligomers to form a soft three-dimensional network. The complete conversion of the reactive liquid to the polymer electrolyte takes around 180 seconds. The final solid polymer electrolyte is a soft material with a glass transition below 230 K and a thermal stability up to 430 K. It may conduct Li$^+$ ions through segmental motion in the network, where the mobility of ions is facilitated by the pendant ethoxy group of the reactive diluent. The whole process, comprising mixing, casting and polymerization, is carried out in an environmentally controlled dry-room (10 m$^2$, R.H.< 2% ±1 at 20 °C) produced by SOIMAR (Caluso, Italy) to avoid the contamination of water molecules and other substances that may influence the performance of the polymeric system.

**Hall effect and double-step chronocoulometry measurements.** Hall effect measurements were performed by supplying a small DC current of a few $\mu A$ at the two current contacts of the patterned devices via a low-noise Keithley 6221 current source and measuring the transverse voltage $V_{xy}$ on two opposite voltage contacts via a low-noise Keithley 2182 nanovoltmeter. The magnetic field was supplied by a calibrated variable-gap magnet made with permanent magnets and able to apply magnetic fields up to 0.7 T at the sample surface. For each gate voltage, $V_{xy}$ was recorded inverting the direction of both the current and the magnetic field and then averaged in order to eliminate any background contribution. Chronocoulometry measurements were performed by applying and then removing a potential step perturbation between the platinum wire (gate electrode) and the negative current contact of the device (working electrode). The small gate current flowing through the electrolyte was measured between the same contacts. The potential step was removed after the current had reached less than one percent of its peak value. Both voltage sourcing and current measurements were performed together via a low-noise Keithley 2400 source measure unit. The measured charge and discharge current curves were fitted separately according to the formulas in Ref. 36. Both Hall effect and chronocoulometry measurements were performed at ambient pressure and room temperature conditions.

**Low-temperature transport experiments.** Temperature-dependent sheet resistance measurements were performed by supplying a small DC current of a few $\mu A$ at the two current contacts of the patterned devices via a low-noise Keithley 6221 current source and measuring the longitudinal voltage $V_{xx}$ on two adjacent voltage contacts via a low-noise Keithley 2182 nanovoltmeter. Each voltage value was measured with both directions of the sourced current and then averaged in order to eliminate thermoelectric voltages present along the voltage leads. Gate voltage was applied at room temperature via a low-noise Keithley 2400 source measure unit and kept constant throughout the entire measurement. Samples were hosted on the cold finger of a Cryomech ST-403 3K pulse tube cryocooler which also provided the temperature control; the cryostat chamber was evacuated to high vacuum in order to prevent vapor liquefaction on the samples during cool down via an Agilent Mini-TASK Turbo Pumping System. Thermal cycles were performed at the slowest obtainable speed in order to minimize the possibility of thermal shock to the samples, i.e., cool down was performed at the maximum power of the in-built Joule heater, while heat up was performed only through the residual thermal conductance of the chamber to the outside environment. The heat up process up to room temperature usually takes more than ten hours during which more than ten thousand resistance values are recorded.

## Acknowledgments

The authors acknowledge the support from the EU Graphene Flagship. Many thanks are due to D. Daghero and G. Profeta for perusing the manuscript and to M. Calandra and F. Dolcini for the useful comments and suggestions on the interpretation of experimental data. We are grateful to E. Cappelluti and S. Galasso for tight-binding and ab-initio DFT calculations.

## Author contributions

R. S. G. conceived and planned the experiments. M. B., S. B. and E. P. realized the FLG devices. J. R. N. and C. G. formulate the PES composition and prepared it. F. P., A. S., M. T. and M. B. performed the transport measurements. E. P., R. S. G. and K. S. analyzed the experimental results. R. S. G. and E. P. wrote the manuscript.

## Additional information

### Competing financial interests

The authors declare no competing financial interests.

**Figures**

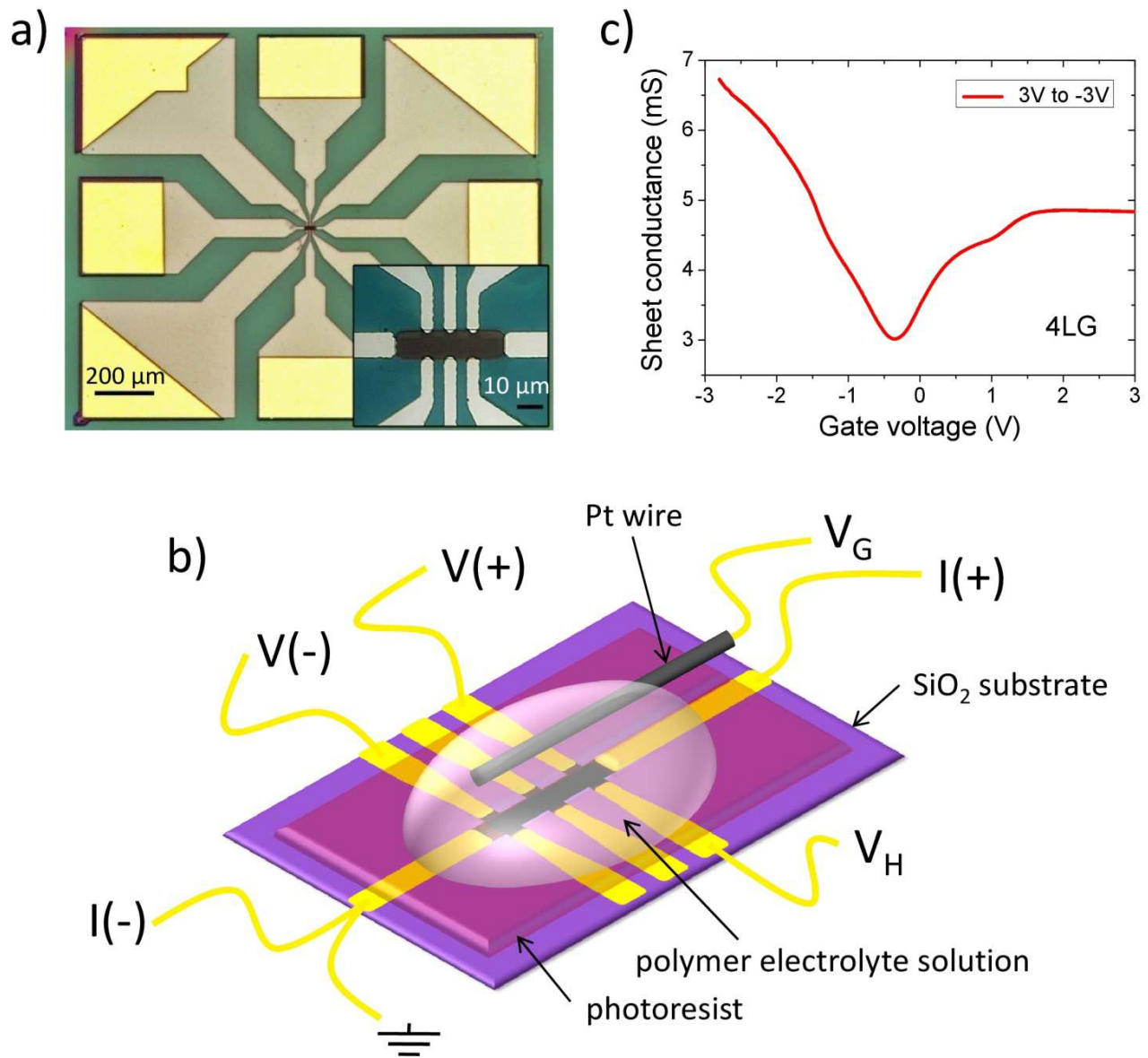

Figure 1 | a) Micrograph of one of the FLG devices where the exposed pads for wire bonding and the photoresist (PR) protection layer are clearly seen. The inset shows a closer view into the region of the Hall-bar shaped FLG flake and the PR window; b) Schematic of the complete device with the PES drop casted over the PR window and the electrical connections for the 4-wire sheet resistance and Hall voltage measurements. The gate voltage $V_G$ is applied between a platinum wire immersed in the PES and the ground electrode; c) An example of a Dirac curve, i.e. a sheet conductance vs. $V_G$ curve recorded between +3 V and -3 V in a 4LG device.

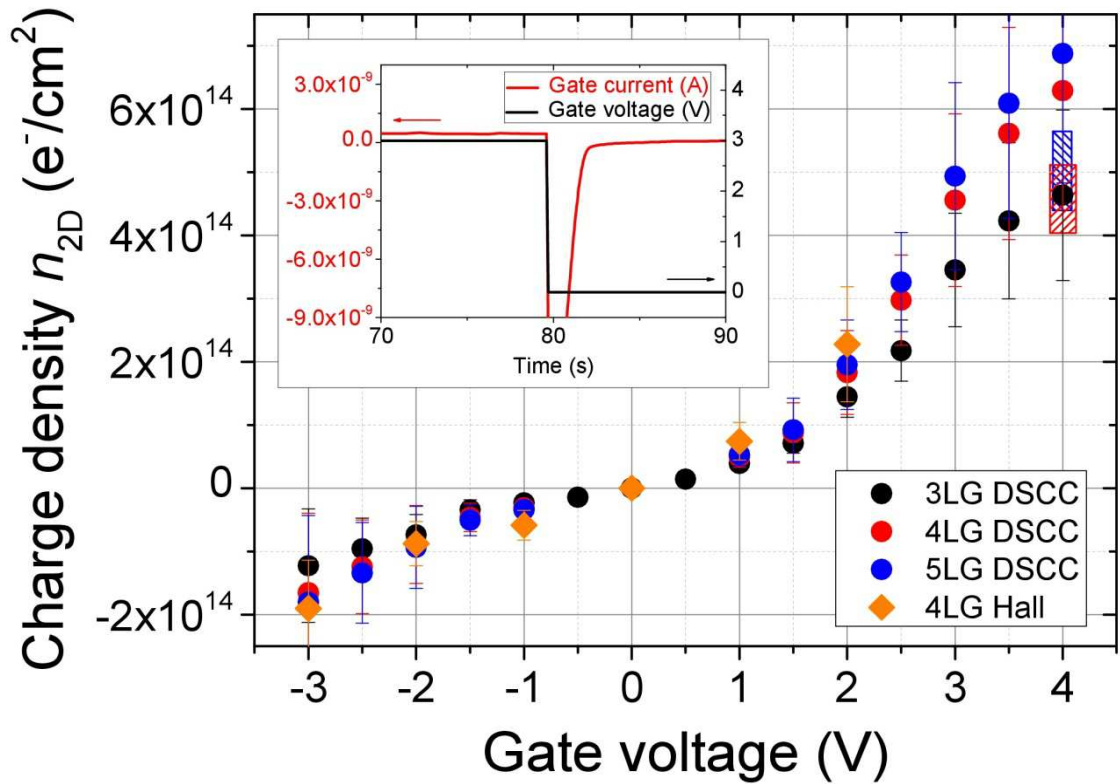

Figure 2 | The induced charge density $n_{2D}$ as function of the gate voltage measured by different techniques in 3LG, 4LG and 5LG devices. The results of Hall effect (orange diamonds) and double-step chronocoulometry (DSCC) measurements (black, red and blue circles) are compared. The range of $n_{2D}$ values estimated at $V_G = 4$V for 4LG and 5LG from an ab-initio evaluation of the quantum capacitance of the devices is also shown (red and blue hatched regions). The inset shows an example of the measured gate current (red line) at the step removal of gate voltage (black line) in a 4LG device. This is a part of the procedure for the determination of $n_{2D}$ by DSCC (see text for details).

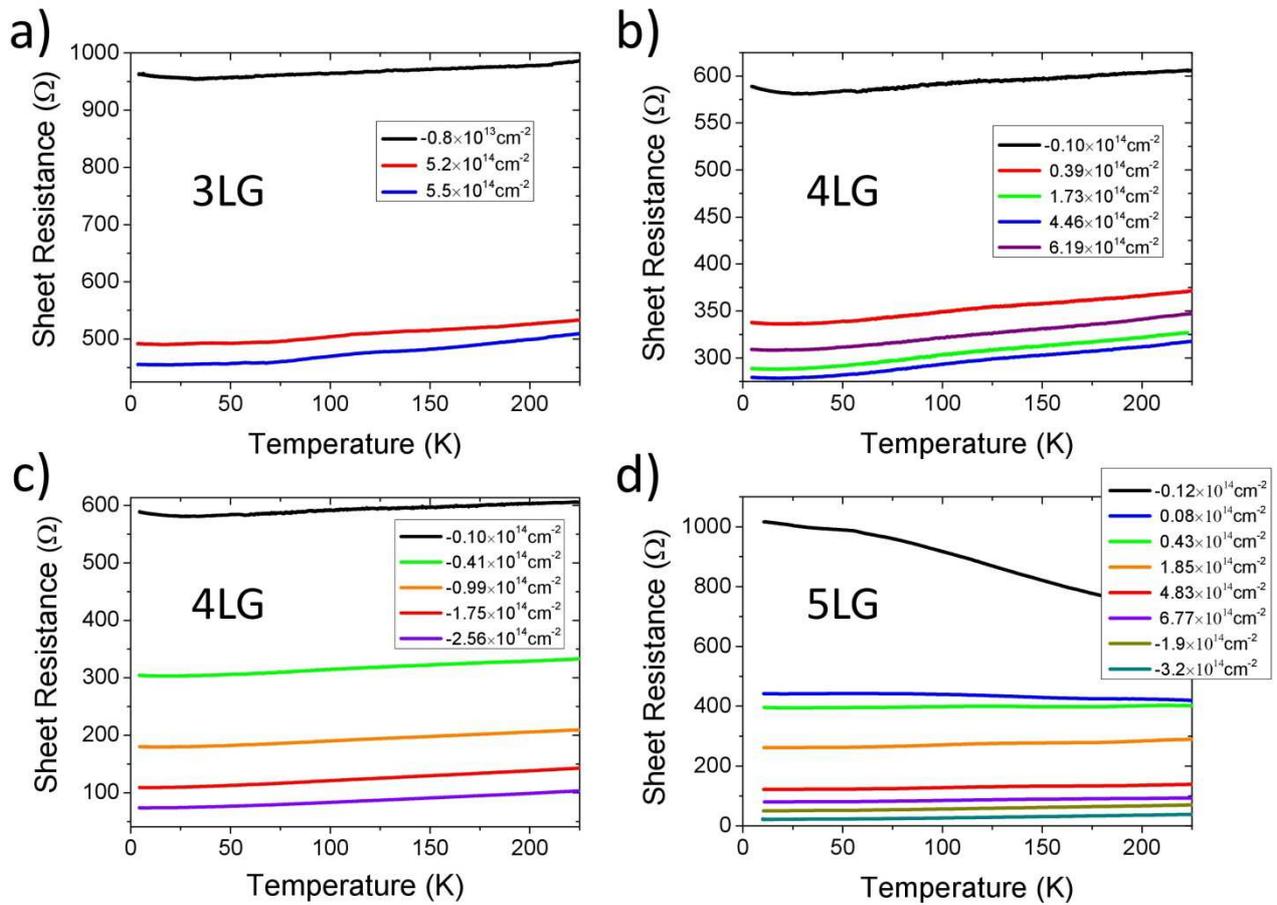

Figure 3 | The experimental sheet resistances as function of temperature measured at different gating-induced surface charge densities (reported in the legends) in a) a 3LG device for positive gating (electron doping); b) a 4LG device for positive gating; c) the same as in b) but for negative gating (hole doping); d) a 5LG device for positive and negative gating.

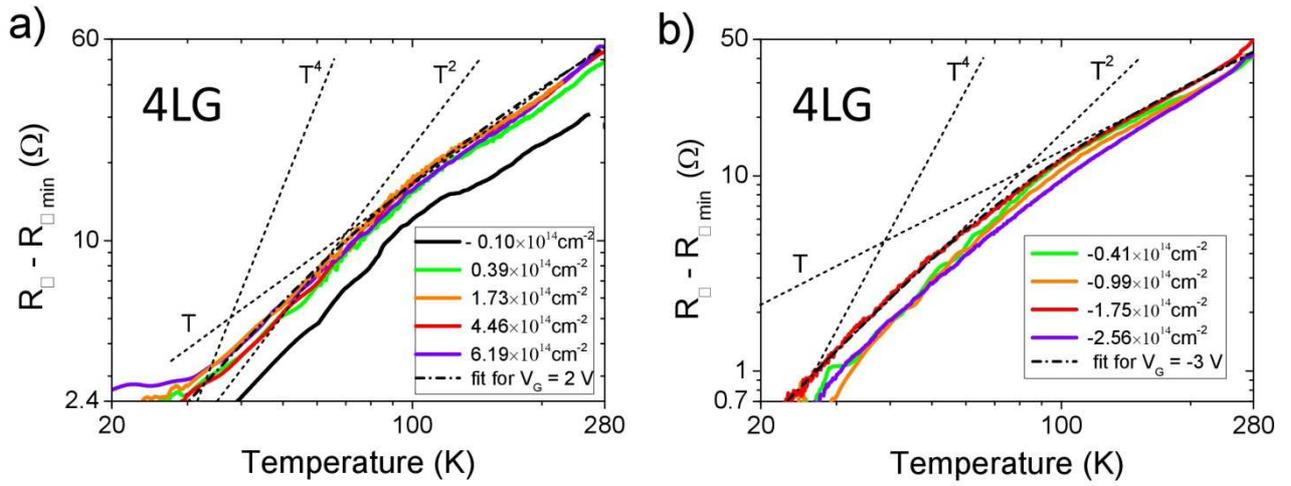

Figure 4 |  Log-log plot of the increase of the sheet resistance in the range 20-280 K for a 4LG device at different surface charge densities induced by a) positive gating (electron doping) and b) negative gating (hole doping). Dotted lines that represent the linear, quadratic and quartic T dependences are also shown for comparison. Dash-dotted lines are the generalized Bloch-Grüneisen fits of the curves at (a) $V_G = 2$ V, i.e. $n_{2D} = 1.73 \cdot 10^{14}$ cm$^{-2}$ and (b) $V_G = -3$ V, i.e. $n_{2D} = -1.75 \cdot 10^{14}$ cm$^{-2}$.

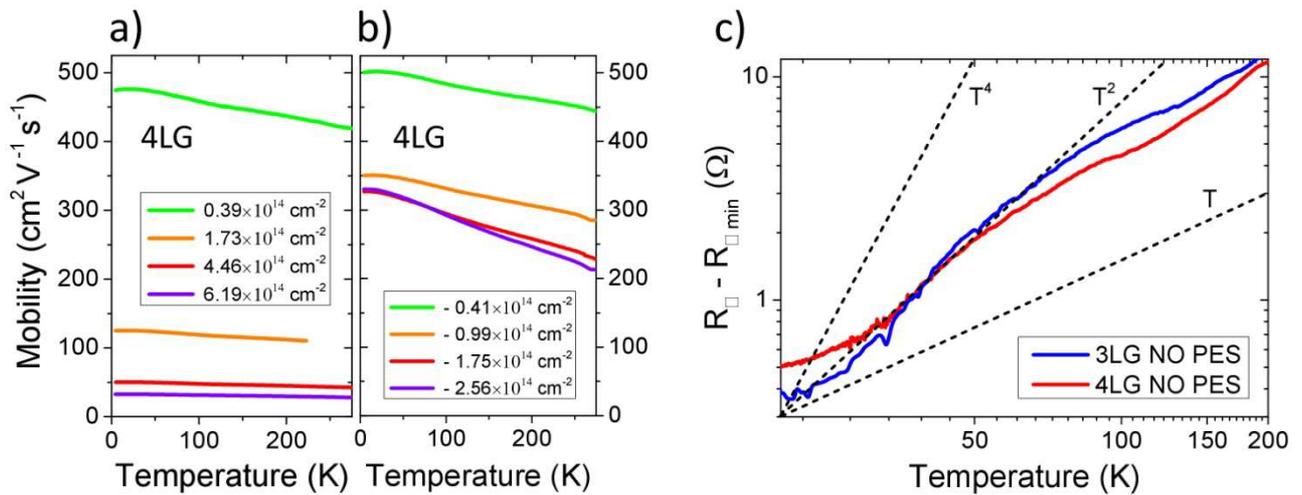

Figure 5 | Carrier mobility as a function of temperature in a 4LG device under electron (a) and hole (b) doping regimes. The mobility is suppressed by an increase in both charge density and temperature. A crossover in the temperature dependence of the mobility is evident around 100 K in all the curves; (c) Longitudinal resistance as a function of temperature for 3LG and 4LG devices before PES drop-casting. Even if the PES is not over the channel we can see the same crossover from a linear to a quadratic temperature dependence around 100 K observed in the gated devices.

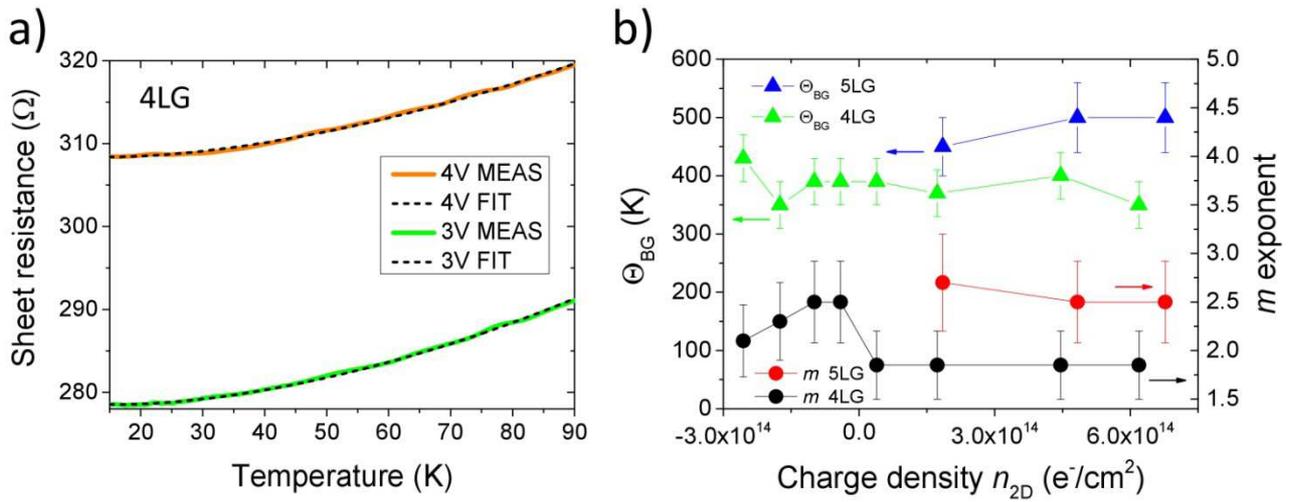

Figure 6 | (a) Longitudinal sheet resistance as a function of temperature between 15 and 90 K for the 4LG device at two different gating levels corresponding to $V_G = 3$ V (green line) and $V_G = 4$ V (orange line). Black dashed lines are the best fits obtained by using the fitting function: $R_{\square\,min} + aT + bT^2$ (see text). They show a remarkable agreement with the experimental data; (b) The Bloch-Grüneisen temperature $\Theta_{BG}$ and the $m$ exponent of the Bloch-Grüneisen model (eq. 3) versus the charge density $n_{2D}$ for 4LG and 5LG devices. While the average values of $\Theta_{BG}$ in 4LG and 5LG are remarkably different, no significant doping dependence of $\Theta_{BG}$ can be appreciated in these curves.